\documentstyle[12pt]{article}
\newcommand{\be}{\begin{equation}} \newcommand{\ee}{\end{equation}}
\newcommand{\bea}{\begin{eqnarray}}\newcommand{\eea}{\end{eqnarray}}

\begin{document}
\begin{titlepage}
%\begin{flushright}
%IP/BBSR/95-105 \\
%hep-ph/9511394 \\
%\end{flushright}
\title { A Heavy Glueball with Color-Singletness Restriction at
Finite Temperature}
\vskip 3.0 cm
\author{ {\bf Lina Paria}$^{1}$, {\bf M. G. Mustafa}$^{2}$ and
{ \bf Afsar Abbas}$^{1}$  \\
 $ ^{1}$Institute of Physics, Sachivalaya Marg, \\
Bhubaneswar-751005, India. \\
$ ^{2} $Variable Energy Cyclotron Centre, 1/AF Bidhan Nagar \\
Calcutta - 700 064. India.}
\footnotetext[1]{e-mail: \ lina/afsar@iopb.ernet.in}
\footnotetext[2]{e-mail: \ mustafa@veccal.ernet.in}
\maketitle
\thispagestyle{empty}
\begin{abstract}
We show that a heavy glueball (much heavier than that studied by
others which is in the range of 1-2 GeV) is generated
in a pure gluon plasma when
color-singletness condition is imposed on the partition function at
finite temperature. This confirms Abbas's recent prediction
(hep-ph/9504430) of the existence of a heavy glueball within
the framework of the early universe scenario.

\end{abstract}
\end{titlepage}
\eject

\eject

\newpage

The study of the strong interaction physics within the farmework of
Quantum Chromodynamics(QCD) [1] clearly indicates the existence
of bound states of gluons --- the so called glueballs.
Much work has been done on glueballs, but all
the theoretical calculations [2,3] estimate the glueballs within the mass
range of a few GeV. Recently Abbas  [4] has shown that the idea of
two distinct phase transition temperatures [5] in QGP can produce very
massive glueballs ( $\ge 45 GeV$ ) --- which he suggested as the
new dark matter (DM) candidate. Here we wish to look at his prediction
of the existence of a heavy glueball in another framework.

Some work has been done [6-11] by imposing color-singletness restriction
on the total partition function of the quark - gluon system to obtain
the thermodynamic quantities and the consequence of the $SU(3)$ color-
singletness restriction on the quark - gluon deconfinement phase diagram
was discussed [8] arround the critical temperature $ T_c $.
 It was pointed out that the color-singletness
constraint gives a finite volume correction automatically.
Auberson et al. [6] and Savatier [7] have suggested a formalism to
project out the allowed color-singlet states of the $ SU(N)$ color group
in the partition function of the quark - gluon system.

Our aim in this letter is to discuss the importance of the $SU(3)$ color-
singletness restriction on the pure gluonic system (glueball).
We will show that if we do not constrain the gluon partition function
to be color-singlet then there exists only one solution for
a glueball bag radius upto a certain temperature (say $T_s$).
 Above $ T_s $ there is no valid solution. But imposing the
color-singletness restriction on the partition function
of the gluonic system, one finds that for $ T < T_s $ we have only
one valid solution for the glueball. For $ T > T_c $ there are
no valid solutions.
Interestingly for $T_s \ < \ T \ < \ T_c$ we obtain two solutions
for the radius of the glueballs. One of the solution gives a very massive
metastable state. This arises due to the finite volume correction
of the color-singletness restriction. Hence we are cofirming Abbas's
recent prediction [4] for the existence of a heavy
glueball arising in cosmological considerations.

The grand canonical partition function [6-11] for a quark-gluon
system is given by

\be
{\cal Z}(\beta ,V) = {Tr} \ \left ( {\hat {\cal P}} \ e^{-\beta \hat
H} \right ) \ ,
\ee

\noindent where $\beta \ = \ {1\over T}$ is the inverse of the temperature,
$V$ is the volume and $\hat H$ is the total Hamiltonian of the physical
system. $\hat {\cal P}$ is the projection operator to give a particular
configuration ( e.g. Angular momentum, Number, Parity, Color etc.)
allowed by the system.

The projection operator for a particular $ j^{th} $
irreducible representation
of a symmetry group $ \cal G $ can be written as [12]

\be
{\hat{\cal P}}_j \ = \ d_j \ \int_{\cal G} d \mu (g) \ \chi^\star_j(g)
\ {\hat U}(g) \ , \
\ee
\noindent
where  $ {\hat U}(g)$is the unitary representation of the group
$ \cal G$ in the Hilbert Space $ \cal H $, $d_j$ and $\chi_j$ ,
are the dimension and the
character of the irreducible representation respectively,
$ d {\mu}(g)$ is the normalized Haar measure in ${\cal G} $.

We will derive the color-singlet partition function for the gluonic
system with $ SU(3) $ color symmetry group. So for $ SU(3) $ color - singlet
configuration $d_j \ = \ 1$ and hence $\chi_j \ = \ 1$.
Since the Hilbert Space is spanned by the gluons only, so
$ {\cal H} \ = \ {\cal H}_{g} $ and the partition function for the
gluonic system becomes (for details see ref.[6-8]),
\be
{\cal Z}(\beta, V) \ = \ \int_{SU(3)} \ d\mu (g) \ Tr \left ( \ {\hat
U_{g}}(g) \ e^{-\beta {\hat H}_{g}} \right ) \ . \
\ee

Where ${\hat H}_{g} $ is the total hamiltonian for gluons.\\
The normalised Haar measure for $ SU(3) $ color group is given by [6,7]

\be
\int  d\mu (g) \ = \ {1 \over {24 \pi^2}} \ \int_{- \pi}^{\pi} \
d \theta_{1} \ d \theta_{2} \ \prod_{i<j}^3 \ {\Big (}2 \ Sin {1 \over 2}
(\theta_i - \theta_j) {\Big )}^2
\ee

Performing the trace operation [6-8] in equation(3) the partition
function becomes,

\be
{\cal Z}(\beta, V) \ = \  \int_{SU(3)}  d\mu(g) \ e^{\Theta} \
,
\ee
\noindent where
\bea
\Theta \ &=& \ \sum_{\alpha} \ \sum^{\infty}_{k=1} \ \Big [
 \ {1\over k} \ \chi_{adj}(g^k) \ e^{-k \beta
\epsilon^{\alpha}_{g}} \ \Big ] \ ,
\eea
\noindent
 and $\chi_{adj}(g^k) \ = \
\ 2 \ + \  \sum^3_{i<j} \ \cos \ k(\theta_i \ - \ \theta_j)$  is
the character of the
adjoint representation of the symmetry  $ SU(3) $ color group related to
 the class paarmeter $\theta_i$  such that \\
 $\sum^3_i \ \theta_i \ = \ 0 \ \ ( mod \  2\pi ) $.

The summation over  the index $k$ arises due to the trace operation
whereas over the index $\alpha$ is due to the discrete single particle
states of gluons with energy $ \epsilon^{\alpha}_{g} $. \\
In a reasonably dense system we can replace $ \sum_\alpha $
 by  $   \ 2 \ \int \ \rho (\epsilon_{g})
\ d\epsilon_{g}  $ in Eq.(6) in the large volume limit.
Here $ \rho ( \epsilon_{g} ) \ = \ {V{\epsilon_{g}}^2 \over {2\pi^2}} $
is the gluon single-particle density of states
 with  $ V $ being the volume and  the factor $ 2 $
in the integration is understood for gluons.\\
After doing the integration in Eq.(6), we get

\be
\Theta \ = \
 {2\pi^2 V\over {45\beta^3}} \ + \ {4V\over
{\pi^2\beta^3}} \ \sum^{3}_{i<j} \ u(\theta_i \ - \ \theta_j) \ ,
\ee
\noindent
where,
\bea
u(\theta_i \ - \ \theta_j) & = & \sum^{\infty}_{k=1}
{\cos \ k (\theta_i \ - \ \theta_j)\over k^4}  \nonumber \\
& = &{\pi^4\over{90}} - {\pi^2\over {12}}(\theta_i \ - \ \theta_j )^2
{ {\Big \{} 1- {|\theta_i \ - \ \theta_j |\over {2\pi}} {\Big \}} }^2
\eea
\noindent
such that
$ (|\theta_i \ - \ \theta_j| < 2\pi) $.
Now, using this expression for $ \Theta $  and integrating the
group integration in  the  saddle point
approximation [6-8], from Eq.(5) finally we get the color-singlet
partition function for the gluons as
\be
{\cal Z}(\beta,V) \  =  \ {\sqrt{3} \over
{3\pi}}  \ \Big [ \ {2V \over { \beta^3}}  \ \Big ]^{-4}  \
\exp \ \Big  [ \ {8{\pi^2}V \over {45 \beta^3}} \ \Big ] \
\ee
%\noindent

This color-singlet partition function for gluons is used to describe the
thermodynamic properties of the glueball in a bag like picture.

Now to understand the effect of the finite volume
correction generated by color-singletness constraint on the partition
function, we consider the gluons as confined in a bag of
radius $ R $ with a pressure constant $ B $. On including the zero
temperature bag term energy [13] $(BV \ + \ C/R)$ we write respectively
the free energy and the energy of the color-singlet glueball in a bag
as,
\bea
F(T,V) \ & = & \ - \ T \ \ln {\cal Z}(T,V) \ + \ BV \ + \ {C \over R}
   \nonumber \\
 & = &  BV \ + \ {C\over R} \ - \ KVT^4 \nonumber \\
& + &  4T\ln \Big [ 2VT^3  \Big ] \ + \ T\ln \Big [\sqrt{3}\pi \Big ] \
\eea

and
\bea
E(T,V) \ & = & \ T^2 \ {\partial\over {\partial T}} \Big ( \ln
{\cal Z}(T,V) \Big) \ + \ BV \ + \ {C \over R} \nonumber \\
&  = &  BV \ + \ {C\over R} \ + \ 3KVT^4 \ - \ 12T \
\eea
\noindent

where
\be
K \ = \ {8 \pi^2 \over {45}}
\ee
\noindent
and $C$ is a positive constant  which depends on
the internal quantum numbers of the bag.

For the stability of the bag[13], the pressure balance condition
 $P \ = \ - \left ({\partial F\over
{\partial V}}\right )_T \ = \ 0$ gives

\be
P \ = \ {C\over {4\pi R^4}} \ - \ B \ + \ KT^4
 \ - \ {3T\over {\pi R^3}} \ = \ 0 \ .
\ee

It is to be noted that the last term in l.h.s. of Eq.(13) is the finite
volume correction due to color-singletness restriction. \\
Using Eq.(13) and eleminating the $ T $ dependence of Eq.(11), we get
the average bag internal energy corresponding to the equilibrium
state as,

\be
E \ = \ 4BV
\ee
Which satisfies the virial theorem.

Now, with $ x \ = \ {1 \over {R}} $, the equation (13) becomes,
\be
x^4 \ = \ ax^3 \ + \ g \ \ ,
\ee

\be where \ \ \ \  a \ = \ {12T\over C}, \ \ \ g \ = \ {4\pi K\over
C}\left (T^4_s \ - \ T^4 \right ) \ \ and \ \ T_s \ = \ { \Big (}{B\over
K} { \Big )}^{1\over 4}
\ee
\noindent

We will see that this $T_s$ behaves as a transition temperature at
which two bag solutions appears indicating a metastable state for a
glueball with heavy mass. These two solutions arises due to
 the $ SU(3) $ color-singletness restriction.

Now, we solve Eq.(15) graphically.
{}From the intersection of the two curves $ y = x^4 $ and
$ y = ax^3 + g $  (as in Fig.1.) we get two solutions (solid circles)
of Eq.(15) for a particular temperature between $ T_s \le T < T_c$ and
 only one solution for $ T < T_s $ and at $ T = T_c $.
In Fig.1. the graphical solution of Eq.(15) is shown for
four different temperatures $ T =210 $ MeV $( < T_s )$,
 $ T = T_s  = 217.218 $ MeV, $ T = 240$ MeV and $ T = T_c =253.3$ MeV
when $ B^{1 \over 4} =250 $ MeV.  The unphysical bag solutions
corresponding to negative $ x $ are excluded.\\

(i) For $ T < T_s $, there is only one solution of the bag (curve (b)
as in Fig.1.) corresponds to a stable glueball with very low mass.

(ii) When $ T = T_s $, $ g = 0 $. \\
Two curves (curve (a)) $ y = x^4 $ and (curve (c)) $ y = a x^3 + g $
intersects at $ x = 0 $ giving  a solution at $ R = \infty $
and at $ x = {12 T_s \over C }$
leading to a finite bag radius $ R = {C \over {12T_s}} $.
Note that here the large mass glueball just starts forming and appears
at infinity.

(iii) When $ T \ > T_s $, the two solutions approaches each other ( see the
intersection of curve (a) and curve (d) of Fig.1.) with one solution for
smaller $ R $ and  the other with larger $ R $ . This indicates
that at $ T_s $, transition occurs from a small stable
bag to a larger metastable bag of glueball. The numerical values of the
solutions shall be discussed shortly. \\

(iv) As the two solutions approaches each other with increasing
temperature, so there must be a critical temperature $ ( T_c) $ at
which both  the solutions meet at one critical point ( see curve (a)
and  curve (e) in Fig.1.), where two curves (a) and (e)
touch tangentially with the same
slope as $ x(T_c) = {9T_c \over C} $. Hence that the critical radius of the
bag of the gluons is $ R(T_c) = {C \over {9T_c}} $ above which no bag
solutions exists. This critical temperature can be obtained from
Eq.(15) as
\be
 T_c \ = \ {T_s \over   \ \Big [ \ 1 \ - \ {2187 \over {4\pi KC^3}}
\ \Big ] \ ^{1/4}}
\ee
provided $ C \ > \ (2187/4\pi K)^{1/3} $.  \\
Here it is to be noted that at equilibrium, one can calculate energy of
the system either from Eq.(11) or Eq.(14) corresponding to different
temperature $ T $.

If we ignore the color-singletness restriction on the partition function,
one obtains the total energy and the free energy of the gluon bag as
respectively,
\be
 E \ = \ BV \ + \ {C\over R} \ + \ 3KVT^4
\ee \nonumber \\
\be
 F \ = \ BV \ + \ {C\over R} \ - \ KVT^4
\ee

So the pressure balance condition gives the expression for $ R $ as
\be
R \ = \ \Big ( \ {C \over {4\pi K}} \ \Big ) \ ^{1 \over 4} \
 \Big ( \ {T_s^4 \ - \ T^4} \ \Big ) \ ^{-{1 \over 4}}
\ee

{}From this equation we see that for $ T < T_s $ , there is only one finite
solution for a bag. For $ T \ = \ T_s $, the bag solution is at
 $ R = \infty $ and for $ T > T_s $ there is no physically real
solution of the bag. So the appearence of large metastable glueball is
because of the $ SU(3) $ color-singletness restriction in the pure gluonic
plasma.

Now we give numerical values for a particular case (see Table 1.).
 For $ B^{1 \over 4} = 250 $ MeV , we get the temperature
$ T_s  = 217.218 $ MeV  from Eq.(16)
and the critical temperature $ T_c = 253.3$ MeV  from Eq.(17).
For a temperature say $ T = 210 $ MeV ($<T_s$) , we get only one real
solution  ( $R = 0.45$ fm corresponding to the intersection of the
curves (a) and (b) in Fig.1.)  of the gluon bag with mass $0.79$ GeV
by solving Eq.(15). As $ T $ exceeds  $ T_s $, we get two
real solutions for the bag, one with a lower radius and the other
with higher radius as shown in Fig.2. For different temperatures between
$ T_s < T < T_c $  ( $218 -250$ MeV) , the low mass state and the high
mass state with different $ R $ values are shown in Table 1. We notice that
the lower radius solution is gradually increasing when the larger
radius solution decreases fast with increasing temperature. At last
at $ T = T_c  = 253.3 $ MeV, we get only one solution with $ R \sim 0.5$ fm
and $ M \sim 1 $ GeV . \\
Similarly we saw that for $ B^{1\over 4} = 400 $ MeV ,
at $ T = 348 $ MeV
 ( $ >T_s = 347.55 $ MeV ) , the massive state  ($ 952 $ GeV)  having
 larger radius  ( $ R = 2.57 $ fm) and the low mass state
 ( $ M = 1.27 $  GeV )  has a radius $ R = 0.28 $ fm .
At $ T = T_c = 405.282 $ MeV, we get only one solution
with $ R = 0.32 $ fm  and $ M = 1.9 $ GeV  and at $ T = 355 $ MeV, the
higher mass state of $ 45 $ GeV  has radius $ 0.93$ fm .

Hence by studying the color projected and the unprojected partition
functions of the gluonic system as  well as their thermodynamic quantities
we see that, the bag radius of the system diverges with temperature
upto $ T = T_s $ for the color unprojected case. Whereas the finite volume
correction to the bag is automatically generated due to the $ SU(3)$
color-singletness restriction at the partition function level.
This causes a metastable state of a heavy glueball within the
temperature range  $T_s \ < \ T \ < \ T_c$ . A heavy glueball
is what Abbas[4] has obtained in his study of the Dark Matter
problem in the early Universe scenario. We have therefore
confirmed his prediction here.
Quite clearly we expect our calculations to have implications for
the early universe QCD phase transition and the QGP phase transition.
These problems are currently under investigation by us.

%\vfill
%\eject

\vfill
\newpage
\begin{figure}

\noindent {\centerline{\bf FIGURE CAPTIONS}} \\
\vskip 1.0 true cm

\caption{ The graphical solutions of Eq.(15) at various temperature
is shown. The curve labelled by {\bf(a)} is the temperature independent
plot of $y=x^4$ vs $x$ . The other
curves are plots of $y=ax^3+g$ vs $x$ labelled by {\bf (b), (c), (d), (e) }
corresponding to temperature  $ T = 210$ MeV ( $ < T_s$ ),
$T \  =  \ T_s $ ( $217.218$ MeV for
$B^{1/4} \ = \ 250$ MeV ), $ T = 240$ MeV,  and $T_c = 253.3$ MeV
respectively. Intersections of curve {\bf (a)} with others are
denoted by solid circles represent physical solutions. The parameter
value $ C = 6.0$ is used.}

\caption{ The solutions of eq.(15) ( i.e. the bag radius ) as a function of
temperature is shown. Here $ C = 6.0 $, $ B^{1\over 4} = 250$ MeV,
$ T_s= 217.218$ MeV , and $ T_c = 253.3$ MeV .}
\end{figure}
{}~
\vskip 1.5 true cm
\centerline {\bf TABLE CAPTIONS}
\noindent {\bf Table 1} \\
Results for the case $ T_s < T <T_c $ with $ B^{1 \over 4}  = 250 $ Mev,
$ T_s = 217.218$ MeV \\
and $ T_c = 253.3$  MeV
\vfill
\newpage

\begin{table}
\centerline {\bf Table 1}
\vskip 0.2 in
\begin{center}
\begin{tabular}{|c|c|c|c|c|}
\hline
\multicolumn{1}{|c|}{$T$ MeV} &
\multicolumn{1}{|c|}{Lower} &
\multicolumn{1}{|c|}{State} &
\multicolumn{1}{|c|}{Higher} &
\multicolumn{1}{|c|}{State} \\
\hline
 & Radius(fm) & Mass(GeV) & Radius(fm) & Mass(GeV) \\
\hline
 218 & 0.45 & 0.79 & 2.87 & 203  \\
220 & 0.45 & 0.79 & 1.81 & 50.7  \\
230 & 0.45 & 0.79 & 0.98 & 8.0  \\
240 & 0.46 & 0.83 & 0.74 & 3.5  \\
250 & 0.48 & 0.94 & 0.59 & 1.76  \\
\hline
\end{tabular}
\end{center}
\end{table}

\end{document}